\documentstyle[12pt]{article}
\newcommand{\slp}{\raise.15ex\hbox{$/$}\kern-.57em\hbox{$\partial$}}
\newcommand{\sla}{\raise.15ex\hbox{$/$}\kern-.57em\hbox{$a$}}
\newcommand{\slA}{\raise.15ex\hbox{$/$}\kern-.57em\hbox{$A$}}
\newcommand{\slB}{\raise.15ex\hbox{$/$}\kern-.57em\hbox{$B$}}
\newcommand{\slb}{\raise.15ex\hbox{$/$}\kern-.57em\hbox{$b$}}
\newcommand{\slW}{\raise.15ex\hbox{$/$}\kern-.57em\hbox{$W$}}

\newcommand{\be}{\begin{equation}}
\newcommand{\ee}{\end{equation}}
\newcommand{\bear}{\begin{eqnarray}}
\newcommand{\ear}{\end{eqnarray}}

\newcommand{\lvec}{{\buildrel\leftarrow\over\partial}}

\begin{document}
\begin{flushright}
HD--THEP--96--46\\
\end{flushright}
\bigskip
\begin{center}
{\bf\LARGE Hamiltonian Embedding of Self-Dual Model}\\
{\bf\LARGE and Equivalence with}\\
{\bf \LARGE Maxwell-Chern-Simons Theory}\\
\vspace{1cm}
R. Banerjee\footnote[1]{On sabbatical leave from
S.N. Bose National Centre for Basic Sciences, DB-17, Sec 1,
Salt Lake, Calcutta-700064, India}\\
\medskip
Instituto de Fisica, Universidade Federal do Rio de Janeiro\\
CP 68528, 21945-970 Rio de Janeiro, RJ\\
Brazil\\
\bigskip
H. J. Rothe and K. D. Rothe\\
\medskip
Institut  f\"ur Theoretische Physik\\
Universit\"at Heidelberg\\
Philosophenweg 16, D-69120 Heidelberg\\
\end{center}
\vspace{2.0cm}
\begin{abstract}
Following systematically the generalized Hamiltonian
approach of Batalin and Fradkin, we demonstrate the equivalence
of a self-dual model with the Maxwell-Chern-Simons theory
by embedding the former second-class theory into a first-class theory.
\end{abstract}
\newpage
\pagestyle{plain}
\setcounter{page}{1}
\section{Introduction}
It has  long been recognized that the addition of a Chern-Simons
three-form  to a Maxwell term in 2+1 dimensions, will generate a
topological  mass for the ``photon'' \cite{DJT}. Another model which
also describes the propagation of a single massive mode in 2+1 dimensions
is the ``self-dual'' (SD) model originally  proposed in ref. \cite{TPV}.
This suggested a possible equivalence between the (second-class) SD-model
and the (first-class) Maxwell-Chern-Simons (MCS) theory, which has
subsequently been demonstrated on a semi-classical level by Deser and
Jackiw \cite{DJ}. Since then there appeared a number of papers studying
different aspects of this equivalence \cite{GRS,BRR,BR}. However, a study
within the generalized canonical framework of Batalin Fradkin and Tyutin
\cite{BF,BT}, allowing for the systematic
conversion of second-class systems
into first-class  ones, is lacking.

In this paper we demonstrate the above
equivalence between these two models
by embedding the SD-model into
a gauge theory, following systematically the procedure of ref. \cite{BF}.
The first
class constraints and the unitarising Hamiltonian are constructed in
section 2. We then
show in section 3 that the partition
function of the SD-model and MCS-theory
are obtained for different choices of gauge,
thereby establishing the claimed
equivalence as a consequence of the
Fradkin-Vilkovisky theorem \cite{FV}.
Our conclusions are given in section 4.

\section{BF Hamiltonian embedding}
The Lagrangian of the self-dual (SD) model \cite{DJ} is given by
\be\label{2.1}
{\cal L}_{SD}=\frac{1}{2}  f^\mu f_\mu-\frac{1}
{2m}\epsilon_{\mu\lambda\nu}
f^\mu\partial^\lambda f^\nu.\ee
It describes a pure second-class system with
three primary constraints
\bear\label{2.2}
T_0&=&\pi_0\approx 0\nonumber\\
T_i&=&\pi_i+\frac{1}{2m}\epsilon_{ij} f^j\approx 0\ear
and a secondary constraint
\be\label{2.3}
T_3=\frac{1}{m}\left( f^0+\frac{1}{m}
\epsilon_{ij}\partial_i f^j\right)
\approx 0,\ee
where $\pi_\mu$ are the momenta canonically
conjugate to $f^\mu$. The
secondary  constraint follows from the requirement
\be\label{2.4}
\{ T_0(x),H_T\}\approx 0,\ee
where $H_T$ is the total Hamiltonian \cite{Di}
\be\label{2.5}
H_T=H_c+\int d^2 x(v^0 T_0+\sum^2_{i=1} v^iT_i)\ee
with $v^0,v^i$ Lagrange multipliers,
and $H_c$ the canonical Hamiltonian
\be\label{2.6}
H_c=\int d^2x\left\{-\frac{1}{2} f^\mu f_\mu +\frac{1}{m}
\epsilon_{ij} f^0 \partial^i f^j\right\}.\ee
The non-vanishing Poisson brackets of the
constraints are given by
\bear\label{2.7}
\{T_0(x),T_3(y)\}&=&-\frac{1}{m}\delta(\vec x-\vec y)\nonumber\\
\{ T_i(x),T_j(y)\}&=&\frac{1}{m}\epsilon_{ij}\delta
(\vec x-\vec y)\nonumber\\
\{T_i(x),T_3(y)\}&=&-\frac{1}{m^2}\epsilon_{ij}\partial_j
\delta(\vec x-\vec y).\ear
In order to establish the connection of the self-dual model with the
Maxwell-Chern-Simons (MCS) gauge theory, we now make use of the formalism
of Batalin, Fradkin and Tyutin \cite{BF,BT} to embed the second-class
SD-model into a first-class  theory.

We begin by converting the second-class constraints into first-class ones.
We closely follow the notation of ref. \cite{BF},
with the commutators in this
reference being obtained from the Poisson brackets via the substitution
$i\{q,p\}\to[q,p]$.
In the notation of \cite{BF} the first-class constraints $T'_\alpha\
(\alpha=0,1,2,3)$ are given by
\be\label{2.8}
T'_\alpha(x)=T_\alpha(x)+\int d^2 y\int d^2z V_\alpha^\gamma
(x,y)\omega_{\gamma\beta}(y,z)\phi^\beta(z)\ee
where $\omega_{\gamma\beta}(x,y)$ and the structure functions $V_\alpha
^\gamma(x,y)$ are restricted by the original second-class constraint
algebra. The tensor $\omega_{\alpha\beta} (x,y)$ is a $c$-numerical,
completely antisymmetric invertible matrix, which in our case has zero
Grassman signature. The field $\phi^\alpha(x)$ satisfies the symplectic
algebra
\be\label{2.9}
\{\phi^\alpha(x),\phi^\beta(y)\}=\omega^{\alpha\beta}(x,y)\ee
where $\omega^{\alpha\beta}(x,y)$ is the inverse of $\omega_{\alpha\beta}
(x,y)$:
\be\label{2.10}
\int d^2 z\omega^{\alpha\gamma}(x,z)\omega_{\gamma\beta}
(z,y)=\delta^\alpha_\beta\delta(\vec x-\vec y).\ee
For the constraint algebra (\ref{2.7}) a possible choice
for $\omega_{\alpha\beta}(x,y)$ and $V_\alpha^\beta(x,y)$
is given by
\bear\label{2.11}
&&\omega_{03}(x,y)=\frac{1}{m}\delta(\vec x-\vec y)\nonumber\\
&&\omega_{ij}(x,y)=\frac{1}{m}\epsilon_{ij}
\delta(\vec x-\vec y)\nonumber\\
&&\omega_{0i}(x,y)=-\frac{1}{m^2}\epsilon_{ik}\partial_k
\delta(\vec x-\vec y)\ear
and
\bear\label{2.12}
&&V_0^3(x,y)=\delta(\vec x-\vec y)\nonumber\\
&&V_i^j(x,y)=\delta_i^j\delta(\vec x-\vec y),\ear
with $V_\alpha^\beta(x,y)=V_\beta^\alpha(y,x)$.
All remaining elements of $V_\alpha^\beta$ vanish.

The inverse $\omega^{\alpha\beta}$ of $\omega_{\alpha\beta}$ is
found from (\ref{2.10}) and (\ref{2.11}) to be
\be\label{2.13}
\omega^{-1}(x,y)=\left(\begin{array}{cccc}
0&0&0&-m\\
0&0&-m&-\partial^x_1\\
0&m&0&-\partial_2^x\\
m&-\partial_1^x&-\partial_2^x&0\end{array}\right)
\delta(\vec x-\vec y)\ee
In terms of the choice (\ref{2.11}) and (\ref{2.12}),
the first-class constraints (\ref{2.8}) read
\bear\label{2.14}
&&T_0'(x)=T_0(x)-\frac{1}{m}\phi^0(x)\nonumber\\
&&T_i'(x)=T_i(x)+\frac{1}{m}\epsilon_{ik}\phi^k(x)-\frac{1}{m^2}
\epsilon_{ik}\partial_k\phi^0(x)\nonumber\\
&&T_3'(x)=T_3(x)+\frac{1}{m}\phi^3(x)-\frac{1}{m^2}
\epsilon_{kl}\partial_l\phi^k(x)\ear
One readily verifies the strongly involutive algebra
\be\label{2.15}
[T_\alpha'(x),T_\beta'(y)]=0.\ee
We next construct the corresponding involutive Hamiltonian
$H'$. Following \cite{BF}, it is given by
\be\label{2.16}
H'=\left\{H[\varphi]\exp\int d^2x\frac{\lvec}{\partial
\varphi^\alpha(x)}\phi^\alpha(x)\right\}_{\varphi=0}\ee
where $\varphi^\alpha(x)$ are real valued fields,
and $H[\varphi]$ is obtained by solving the equation
\be\label{2.17}
\frac{\partial H[\beta\varphi]}{\partial\beta}=(i)^{-2}
[H[\beta\varphi],[\Omega,\int d^2x\bar\Omega_\alpha(x)
\varphi^\alpha(x)]].\ee
subject to the boundary condition $H[0]=H_c$.
The operators $\Omega$ and $\bar\Omega_\alpha$ are defined
in ref. \cite{BF}, and in our
particular case have the simple form
\bear\label{2.18}
&&\Omega=\int d^2xT_\alpha(x){\cal C}^\alpha(x)\nonumber\\
&&\bar\Omega_\alpha(x)=\int d^2y
\bar V_\alpha^\gamma(x,y)\bar{\cal P}_\gamma(y)\ear
where $\bar V_\alpha^\gamma$ is the inverse of
$V_\alpha^\gamma$, as given by
\bear\label{2.19}
&&\bar V_0^3(x,y)=\bar V_3^0(\vec y,\vec x)=\delta(\vec x-\vec y)\nonumber\\
&&\bar V_1^1(x,y)=\bar V_2^2(x,y)=\delta(\vec x-\vec y)\ear
with the remaining matrix elements vanishing. The fields
${\cal C}^\alpha(x)$ and $\bar{\cal P}_\beta(x)$ represents a
canonical ghost pair with opposite Grassman parity
to that of the constraints $T_\alpha$:
\be\label{2.20}
[{\cal C}^\alpha(x),\bar{\cal P}_\beta(y)]=
i\delta^\alpha_\beta\delta
(\vec x-\vec y).\ee
Defining the generating functional
\be\label{2.21}
L[\varphi]=\int d^2xd^2y\varphi^\alpha(x)
V_\alpha^\gamma(x,y)T_\gamma(y)\ee
the solution to (\ref{2.17}) for $\beta=1$ can be written
in the form
\be\label{2.22}
H[\varphi]=e^{iL[\varphi]}H_ce^{-iL[\varphi]}.\ee
Expanding the exponentials in powers of $L[\varphi]$,
one finds that the series truncates after the third term,
with the result
\bear\label{2.23}
H[\varphi]&=&H_c+[iL,H_c]+\frac{1}{2!}[iL,[iL,H_c]]\nonumber\\
&=&H_c-\int d^2z\left\{m\varphi^3(z)T_3(z)+\frac{1}{m}
\varphi^k(z)(\epsilon_{kl}\partial_lf^0(z)-f^k(z))\right\}
\nonumber\\
&&-\frac{1}{2!}\int d^2z\left\{\varphi^3(z)\varphi^3(z)-
\varphi^k(z)\varphi^k(z)-\frac{2}{m}\epsilon_{kl}
\varphi^3(z)\partial_l\varphi^k(z)\right\}\nonumber\\
&&\ear
where $k$ and $l$ take the values 1,2. From here, and the definition
(\ref{2.16}) one obtains
for the involutive Hamiltonian
\bear\label{2.24}
H'&=&H_c-\int d^2z\ m\phi^3(z)T_3'(z)+\frac{1}{2}\int
d^2z(\phi^k(z)\phi^k(z)+\phi^3(z)\phi^3(z))\nonumber\\
&&-\int d^2z\phi^k(z)\left(\frac{1}{m}\epsilon_{kl}
\partial_lf^0(z)-f^k(z)\right),\ear
with the strongly involutive property
\be\label{2.25}
[H',T_\alpha'(x)]=0.\ee
One readily checks that the involutive Hamiltonian (\ref{2.24})
is obtained from the canonical one by a simple translation in the
fields
\be\label{2.26}
H'=H_c[f^i+\phi^i,f^0+\phi^3].\ee
This property will be useful in the forthcoming analysis.

We next proceed to construct the BRST Hamiltonian and the
corresponding partition function. From the work of ref.
\cite{BF} it is readily seen that
\be\label{2.27}
H'=H_{\rm BRST},\ee
since $H'$ is strongly involutive. The remaining step concerns
the construction of the unitarizing Hamiltonian $H_U$ as
defined in \cite{BF}
\be\label{2.28}
H_U=H_{\rm BRST}+\frac{1}{i}[\Psi,Q]\ee
where the nilpotent BRST charge $Q$ and the fermion
function $\Psi$ are given by
\be\label{2.29}
Q=\int d^2z(T_\alpha'C^\alpha+p_\alpha{\cal P}^\alpha)\ee
\be\label{2.30}
\Psi=\int d^2z(\bar C_\alpha\chi^\alpha+\bar{\cal P}_\alpha
q^\alpha)\ee
and where $(q^\alpha,p_\alpha)$ are canonically conjugate
multiplier fields, and $\chi^\alpha$ are gauge
fixing functions having
the same statistics as the constraints, and are required to satisfy
\be\label{2.31}
det\{\chi^\alpha,T_\beta'\}\not=0\ee
This completes our analysis on the operator level.

We now consider the partition function corresponding
to the unitarizing Hamiltonian (\ref{2.28}):
\be\label{2.32}
Z=\int[D\mu](det\ \omega)^{-1/2}e^{iS_U}\ee
where
\be\label{2.33}
S_U=\int d^3x\Bigl\{\pi^\mu \dot f^\mu+\frac{1}{2}\phi^\alpha
\omega_{\alpha\beta}\dot\phi^\beta+p_\alpha\dot q^\alpha+
{\cal P}_\alpha\dot{\bar C}^\alpha+C_\alpha\dot{\bar{\cal P}}
^\alpha
-H'-\frac{1}{i}[\Psi,Q]\Bigr\}\ee
and the integration measure $[d\mu]$
involves all the fields appearing in the exponent. The second term
in (\ref{2.33}) is due to the symplectic algebra (\ref{2.9})
satisfied by the $\phi$-fields, and the field independence
of the corresponding symplectic metric $\omega^{\alpha\beta}
$ \cite{A}. This term (including $det\ \omega$) can in principle
be put into a standard canonical form via a Darboux transformation \cite{J}.
In the present case it is, however, more convenient.
to proceed directly from (\ref{2.33}).

In the following we consider gauge functions $\chi^\alpha$ which
do not depend on the multipliers $q^\alpha, p_\alpha$. By suitably
rescaling the fields \cite{FV,Hen}
$\chi^\alpha\to\chi^\alpha/\beta,\ p_\alpha\to\beta p_\alpha,\
\bar C_\alpha\to\beta\bar C_\alpha$ (the superjacobian
of this transformation is unity) and taking $\beta\to0$, one
can carry out the integrals over the ghost variables and multiplier
fields, with the result
\be\label{2.34}
Z=\int D\pi_\mu Df^\mu D\phi^\alpha\delta[\chi]
\delta[T']det[\chi^\alpha,T_\beta']
e^{i\int d^3x\{\pi_\mu\dot f^\mu+\frac{1}{2}
\phi^\alpha\omega_{\alpha\beta}\dot\phi^\beta-{\cal H}
_c[f^i+\phi^i,
f^0+\phi^3]\}}\ee
where we have used (\ref{2.26}) and have dropped
$(det\ \omega)^{-1/2}$, which is just an (irrelevant) constant.

\section{Equivalence of SD and MCS model}
\setcounter{equation}{0}
In this section we prove the equivalence of
the self-dual \cite{DJ} and the Maxwell-Chern-Simons theory, by
using the Fradkin-Vilkovisky theorem \cite{FV} stating
the gauge independence of the partition function
(\ref{2.34}). To start out with, we choose a class of gauges
resticted by the two gauge conditions
\be\label{3.1}
\chi^i=\phi^i-\frac{1}{m}\partial_i\phi^0\approx 0,
\qquad i=1,2\ee
The other two conditions $\chi^0\approx\chi^3\approx0$
are still arbitrary. Using (\ref{3.1}), all momentum
integrals as well as $\phi^1, \phi^2$ and $\phi^3$ integrals
can be done trivially, leading to the result
\bear\label{3.2}
&&Z=\int Df^\mu D\phi^0\delta[\chi^0]\delta[\chi^3]
\nonumber\\
&&\times\exp i\int\Bigl\{-\frac{1}{m^2}\phi^0\epsilon_{ij}\partial_i\dot f^j+
\frac{1}{m}f^i\partial_i\phi^0\nonumber\\
&&+\frac{1}{m}f^1\dot f^2-\frac{1}{2m^2}(\epsilon_{ij}
\partial_if^j)^2-\frac{1}{2}\vec f^2-\frac{1}{2m^2}
(\partial_i\phi^0)^2\Bigr\}\ear
where the property $\epsilon_{ij}\partial_i\phi^j=0$, following
from (\ref{3.1}) has been used. Note that at least one of the
gauge conditions $\chi^0\approx\chi^3\approx 0$ must involve
$f^0$ in order to render the $f^0$-integral finite.

\bigskip
\noindent{\it a) Recovery of the SD-model in the unitary gauge}

\bigskip\noindent
Choosing for the remaining gauge conditions
\be\label{3.3}
\chi^0=\phi^0\approx0\ee
\be\label{3.4}
\chi^3=T^3=f^0+\frac{1}{m}\epsilon_{ij}\partial_if^j\approx0\ee
one is led to the partition function of the self-dual model,
\be\label{3.5}
Z=\int Df^\mu\delta[f^0+\frac{1}{m}\epsilon_{ij}\partial_if^j]
e^{i\int{\cal L}_{SD}}\ee
Note that to arrive at this expression, repeated use has been
made of the gauge condition (\ref{3.4}). This gauge condition
is just the original second-class constraint $T^3\approx0$.
Similarly from $\delta[T'^{0}]$ in (\ref{2.34}) we see that
the condition (\ref{3.3}) can
also be viewed as the original second-class constraint
$T^0=\pi^0\approx0$. For this reason the above gauge is referred
to in the literature \cite{BF,BT} as the ``unitary'' gauge.

\pagebreak
\noindent{\it b) Recovery of the MCS model in the Coulomb-like gauge}

\bigskip\noindent
Consider the gauge condition \cite{BR}
\be\label{3.6}
\chi^0, \chi^3:f^i-\frac{1}{m}\epsilon_{ij}\partial_jf^0\approx0\ee
Since this gauge implies $\partial_if^i=0$, we refer to it
as the Coulomb-like gauge. Using (\ref{3.4}), the partition
function (\ref{3.2}) can be put into the form
\bear\label{3.7}
&&Z=\int Df^\mu D\phi^0\delta[f^i-\frac{1}{m}\epsilon_{ij}\partial_j
f^0]\times\nonumber\\
&&\times\exp i\int d^3x\Bigl\{-\frac{1}{m^2}\phi^0\epsilon_{ij}\partial_i
\dot f^j+\frac{1}{2m^2}\phi^0\vec\nabla^2\phi^0\nonumber\\
&&+\frac{1}{m}f^1\dot f^2-\frac{1}{2m^2}(\epsilon_{ij}
\partial_if^j)^2-\frac{1}{2m^2}(\epsilon_{ij}\partial_jf^0)^2\Bigr\}
\ear
Performing the Gaussian $\phi^0$-integration we find
\bear\label{3.8}
Z&=&\int Df^\mu\delta[f^i-\frac{1}{m}\epsilon_{ij}\partial_j
f^0]\times\nonumber\\
&&\exp i\int d^3x\Bigl\{-\frac{1}{2m^2}(\epsilon_{ij}\partial_i\dot f^j)
\frac{1}{\vec\nabla^2}(\epsilon_{kl}\partial_k\dot f^l)\nonumber\\
&&+\frac{1}{2m}\epsilon_{ij}f^i\dot f^j-
\frac{1}{2m^2}(\epsilon_{ij}\partial_if^j)^2-
\frac{1}{2m^2}(\epsilon_{ij}\partial_jf^0)^2\Bigr\}\ear
Making repeated use of the gauge condition, the partition
function can be put into the manifestly covariant form
\be\label{3.9}
Z=\int Df^\mu\delta[f^i-\frac{1}{m}\epsilon_{ij}\partial_j
f^0]e^{i\int d^3x\ {\cal L}_{\rm MCS}}\ee
where ${\cal L}_{\rm MCS}$ is the Maxwell-Chern-Simons Lagrangian
\be\label{3.10}
{\cal L}_{\rm MCS}=-\frac{1}{4m^2}f_{\mu\nu}f^{\mu\nu}-\frac{1}{2m}
\epsilon_{\mu\nu\lambda}f^\mu\partial^\nu f^\lambda\ee
It is simple to see that the gauge condition appearing
in (\ref{3.9}) is equivalent to the Coulomb gauge and
Gauss law in that gauge. This completes the proof of the equivalence
of the self-dual model and the Maxwell-Chern-Simons theory within
the BF framework.

\section{Conclusion}
The equivalence of the self-dual (SD) model of ref. \cite{TPV}
and the Maxwell-Chern-Simons theory has been discussed using
different approaches \cite{DJ,GRS,BRR,BR}.
It was the objective of this
paper to demonstrate this equivalence by suitably extending the
phase space of the SD-model, following systematically
the procedure of ref. \cite{BF}. This involved the conversion
of the four second-class constraints of the SD model into the
first-class ones, and the corresponding construction of the unitarising
Hamiltonian. The gauge-invariant partition function associated
with this Hamiltonian was shown to reduce to that of the
original second-class SD-model in the ``unitary gauge'',
and to the MCS theory in a Coulomb-like gauge. It is interesting
to note that the field $f^\mu$ in the embedded partition function
(\ref{2.34}) played the role of either the fundamental field
of the SD-model or the gauge potential of the MCS theory, in
the unitary and Coulomb-like gauge, respectively.

Finally, we wish to point out that the equivalence investigated
here has proven useful in the study of abelian bosonization in 2+1
dimensions of massive fermionic models \cite{FS}, \cite{B}.

\section{Acknowledgement}
One of the authors (R.B.) would like to thank the CNPq for
financial support.


\begin{thebibliography}{12}
\bibitem{DJT} S. Deser, R. Jackiw, and S. Templeton,
Ann. Phys. {\bf 140} (1982) 372.
\bibitem{TPV} P. K. Townsend, K. Pilch, and
P. Van Nieuwenhuizen, Phys. Lett. {\bf 136B} (1984) 38.
\bibitem{DJ} S. Deser and R. Jackiw, Phys. Lett. {\bf 139B}
(1984) 371.
\bibitem{GRS} R. Gianvittorio, A. Restuccia, and
J. Stephany, Mod. Phys. Lett. {\bf A6} (1991) 2121.
\bibitem{BRR} R. Banerjee, H. J. Rothe, and K. D. Rothe,
Phys. Rev. {\bf D52} (1995) 3750.
\bibitem{BR} R. Banerjee and H. J. Rothe, Nucl. Phys. {\bf B447} (1995) 183.
\bibitem{BF} I. A. Batalin and E. S. Fradkin, Nucl. Phys.
{\bf B279} (1987) 514.
\bibitem{BT} I. A. Batalin and I. V. Tyutin, Int. J. Mod. Phys.
{\bf A6} (1991) 3255.
\bibitem{A} R. Amorim, Z.Phys. {\bf C67} (1995) 695.
\bibitem{J} R. Jackiw, "Constrained Quantization without Tears" ("Diverse
Topics in
Theoretical and Mathematical Physics", World-Scientific, Singapore, 1995.)
\bibitem{FV} E. S. Fradkin and G. Vilkovisky,
Phys. Lett. {\bf B55} 91975) 224.
\bibitem{Hen} M. Henneaux, Phys. Rep. {\bf 126} (1985) 1.
\bibitem{Di} P. A. M. Dirac, ``Lectures on Quantum
Mechanics'', Belfer Graduate School of Science,
Yeshiva University, N.Y. 1964.
\bibitem{FS} E. Fradkin and F. Schaposnik, Phys. Lett. {\bf B338} (1994) 253.
\bibitem{B} R. Banerjee, Phys. Lett. {\bf B358} (1995) 297;
Nucl. Phys. {\bf B465} (1996) 157.
\end{thebibliography}
\end{document}